\def \SAIT #1 #2 {{\em Mem.\ Soc.\ Astron.\ It.\/} {\bf #1}, #2}
\def \MESS #1 #2 {{\em The Messenger\/} {\bf #1}, #2}
\def \ASTRNACH #1 #2 {{\em Astron. Nach.\/} {\bf #1}, #2}
\def \AAP #1 #2 {{\em Astron. Astrophys.\/} {\bf #1}, #2}
\def \AAL #1 #2 {{\em Astron. Astrophys. Lett.\/} {\bf #1}, L#2}
\def \AAR #1 #2 {{\em Astron. Astrophys. Rev.\/} {\bf #1}, #2}
\def \AAS #1 #2 {{\em Astron. Astrophys. Suppl. Ser.\/} {\bf #1}, #2}
\def \AJ #1 #2 {{\em Astron. J.\/} {\bf #1}, #2}
\def \ANNREV #1 #2 {{\em Ann. Rev. Astron. Astrophys.\/} {\bf #1}, #2}
\def \APJ #1 #2 {{\em Astrophys. J.\/} {\bf #1}, #2}
\def \APJL #1 #2 {{\em Astrophys. J. Lett.\/} {\bf #1}, L#2}
\def \APJS #1 #2 {{\em Astrophys. J. Suppl.\/} {\bf #1}, #2}
\def \APSS #1 #2 {{\em Astrophys. Space Sci.\/} {\bf #1}, #2}
\def \ASR #1 #2 {{\em Adv. Space Res.\/} {\bf #1}, #2}
\def \BAIC #1 #2 {{\em Bull. Astron. Inst. Czechosl.\/} {\bf #1}, #2}
\def \JSQRT #1 #2 {{\em J. Quant. Spectrosc. Radiat. Transfer\/} {\bf #1}, #2}
\def \MN #1 #2 {{\em Mon. Not. R. Astr. Soc.\/} {\bf #1}, #2}
\def \MEM #1 #2 {{\em Mem. R. Astr. Soc.\/} {\bf #1}, #2}
\def \PLR #1 #2 {{\em Phys. Lett. Rev.\/} {\bf #1}, #2}
\def \PASJ #1 #2 {{\em Publ. Astron. Soc. Japan\/} {\bf #1}, #2}
\def \PASP #1 #2 {{\em Publ. Astr. Soc. Pacific\/} {\bf #1}, #2}
\def \NAT #1 #2 {{\em Nature\/} {\bf #1}, #2}

\documentstyle{memsait}
\input epsf.sty
\begin{opening}
\title{HST OBSERVATIONS OF FR~I RADIOGALAXIES: WHAT DO THEY 
TELL US ABOUT THE BL LAC - FR~I UNIFICATION SCHEME?} 
\author{MARCO CHIABERGE$^1$, ANNALISA CELOTTI$^1$, ALESSANDRO CAPETTI$^2$,
GABRIELE GHISELLINI$^3$}
\institute{$^1$ SISSA/ISAS, Trieste, Italy \\
$^2$ Osservatorio Astronomico di Torino, Pino Torinese (TO), Italy\\
$^3$ Osservatorio Astronomico di Brera, Merate (LC), Italy}
\date{} 
\end{opening}

\begin{document}

\oddpagefooter{}{}{} 
\evenpagefooter{}{}{} 
\ 
\bigskip

\begin{abstract}
We explore the viability of the unification of BL Lacs and FR~I radio
galaxies by comparing the core emission of radio galaxies  
with those of BL Lacs,
taking advantage of the newly measured optical nuclear
luminosity of FR~I sources. 
The spectral properties of complete samples
are studied in the radio-optical luminosity plane: we calculate the
predicted luminosity of BL Lacs when observed off-axis, 
in the frame of a simple one--zone model.
We find that the bulk Lorentz factors required in order to account for
the observed luminosities are significantly smaller than
those implied by other, both observational and theoretical, considerations.
In order to reconcile these results with the unification scheme,
velocity structures in the jet are suggested, where a fast spine is
surrounded by a slower layer. 

\end{abstract}

\section{Introduction}
Unification models adduce the main differences between the observed
properties of different classes of AGNs to the anisotropy of the
radiation emitted by the active nucleus (see Urry \& Padovani 1995 for a
review). In particular, for low
luminosity radio-loud objects, namely BL Lacs and FR~I radio galaxies, 
it is believed that this effect is mainly due to relativistic beaming.
Within this scenario, the emission from the inner regions of
a relativistic jet dominates the observed radiation in BL Lacs, while
in FR~I, whose jet is
observed at larger angles with respect to the line of sight,
this component is strongly debeamed.  Evidence
for this unification scheme includes the power and morphology of the
extended radio emission of BL Lacs and the properties of their host galaxies, 
which are similar to those of FR~I.
Despite this global agreement, it should be stressed that beaming
factors inferred from the broad band spectral
properties of blazars
are significantly and systematically larger than those 
suggested by radio luminosity data.

Thanks to the Hubble Space Telescope (HST), 
the optical counterparts of the radio cores have been recently 
discovered in FR~I galaxies (Chiaberge et al. 1999). 
In this paper we directly compare the properties
of the optical and radio cores
of FR~I radio galaxies belonging to the 3CR catalog 
with those of two complete samples of 
BL Lacs derived from the 1Jy catalog and from the {\it Einstein} Slew Survey.

\section{Core versus extended luminosity}

We firstly consider the optical and radio core luminosities separately,
comparing the nuclear emission of FR~I and BL Lacs for
bins of equal extended radio power (see Fig 1 of Chiaberge et al. 2000).  
The BL Lacs are at least 4 order of magnitude brighter in the optical 
than FR~I nuclei. As the core radiation of BL Lacs is
enhanced by relativistic beaming, we can derive the bulk Lorentz factors
$\Gamma$ requested to account for the observed spread, 
if typical observing angles are known.
If BL Lacs are observed at an angle $\theta\sim 1/\Gamma$ and $\theta\sim 
60^{\circ}$ is the median observing angle of FR~I galaxies,
bulk velocities with $\Gamma$ $\sim 4-6$ are requested to account 
for the optical enhancement of BL Lacs in each 
bin of extended luminosity. A similar result is obtained by comparing
the radio core emission (e.g. Kollgaard et al. 1996).
However, as already mentioned, such values are significantly and 
systematically lower than those required by other independent means,
such as superluminal motions and high energy spectral constraints
in both LBLs and HBLs. 
These in fact require values of the Doppler factor
$\delta$ ($=\Gamma$ for a viewing angle $\theta=1/\Gamma$)
in the range 15--20 for the region emitting most of the 
radiation in both HBLs and LBLs.

\section{FR~I and BL Lac in the  $L_o$ and $L_r$ plane and jet velocity
structures}

\begin{figure}
\epsfysize=6cm 
\hbox{\hspace{0.5cm}\epsfbox{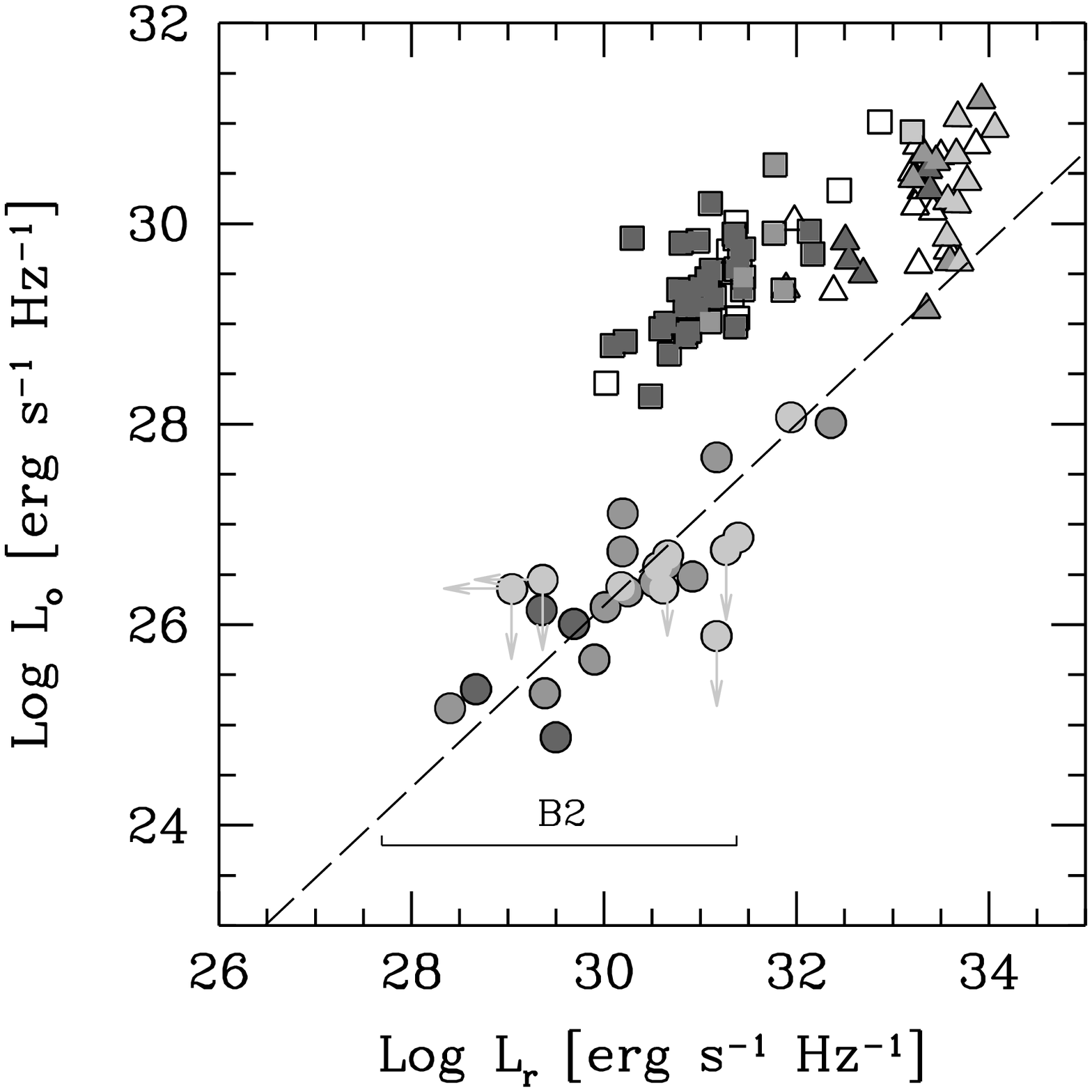} \epsfysize=6cm \epsfbox{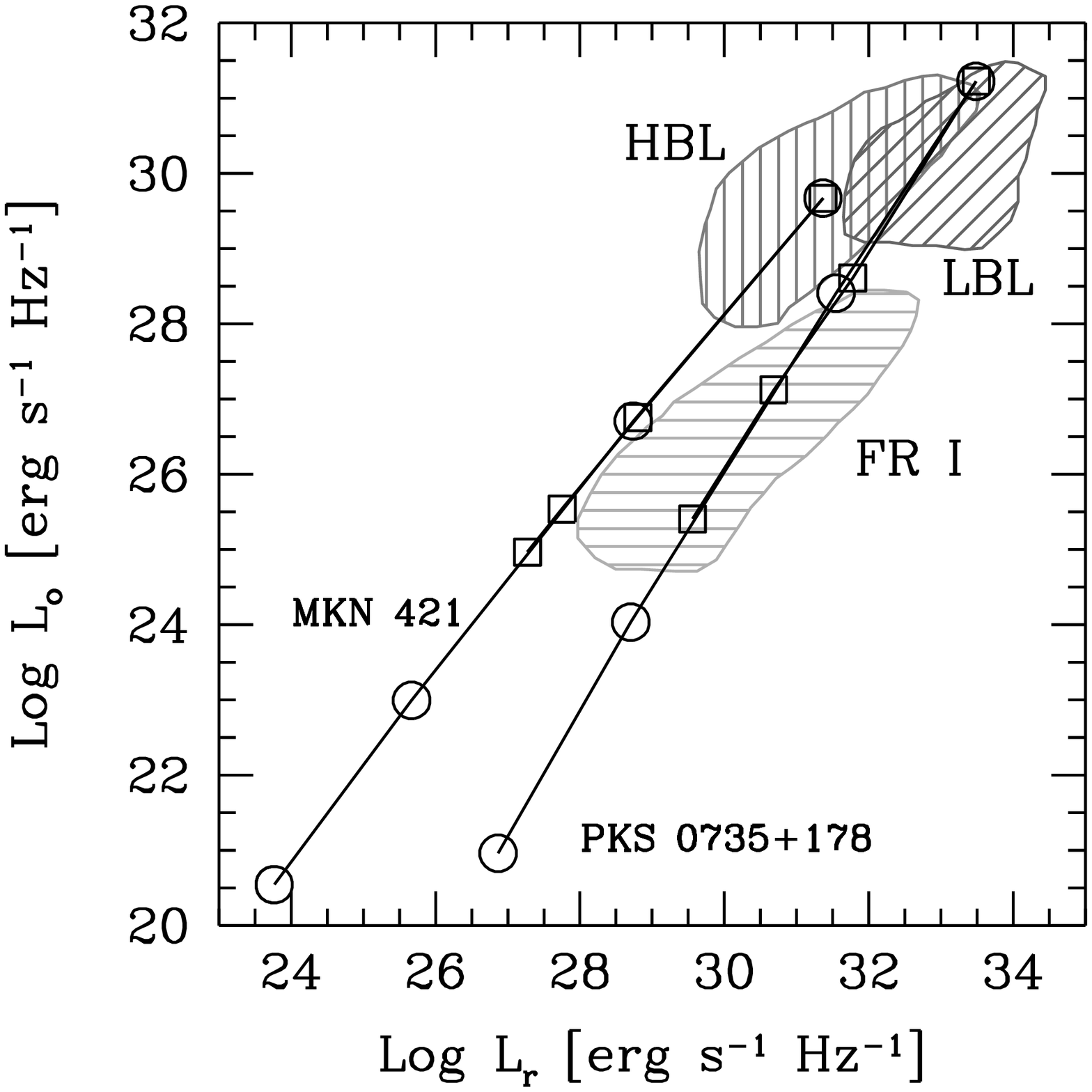}} 
\vspace{-0.5cm}
\caption[h]{{\bf Left (a):} nuclear luminosity of BL Lac (Slew=squares, 
1Jy=triangles) and FR~I (circles) samples in the $L_o-L_r$ plane. 
The greyscale represents three different bins of extended luminosity 
(1.4 GHz), $\log L_e<31.5$ between $31.5$ and $32.5$, and $> 32.5$.
{\bf Right (b):} debeaming trails of two BL Lacs for the one-zone model. 
Circles represent the object as observed at $\theta\sim 1/\Gamma$, 
$10^{\circ}$, $30^{\circ}$ and $60^{\circ}$. Squares represent the
same angles of view but for the two velocity jet.}
\label{fig1}
\end{figure}

In Fig. \ref{fig1}a we show the optical vs radio core luminosity
for the three samples. 
In order to determine how beaming affects the observed luminosities
and thus how objects could be connected in the $L_o-L_r$ plane, we
consider the SED of BL Lacs, observationally much better determined,
and calculate the observed spectrum of the misaligned objects, by
taking into account the relativistic transformations.
In fact an important and previously neglected point, is that these
transformations depend on the spectral index in the band considered,
which in itself might change as a function of the degree of beaming.
Therefore, in order to correctly de--beam the SED of BL Lacs, a
continuous representation of the SED and an estimate of the bulk Lorentz
factor of the emitting region are needed (it is again assumed
$\theta=1/\Gamma$). 
We derive both the continuous description of the SED and
the bulk Lorentz factor of the source by adopting a
homogeneous synchrotron self--Compton emission model.
We model the observed SED of two sources (Mkn~421 and PKS 0735+178), 
representative of High and Low Energy-peaked BL Lacs (HBL and LBL), 
respectively.
We then calculate the corresponding observed SEDs for different orientations.  
Clearly the net effect of debeaming is a ``shift'' of the SED towards 
lower luminosities and energies.
As expected, for $\theta=60^{\circ}$ the resulting ``debeamed'' BL Lac 
component is about 
four orders of magnitude below the radio galaxy region in the optical, 
and two/four in the radio band (see Fig \ref{fig1}b).
While equivalently incompatible with the FR~I population properties, the HBL and
LBL move on different trails. This effect is due to the different shape of
their SED.
The simplest and rather plausible hypothesis to account for this
discrepancy within the unification scenario is to assume a structure
in the jet velocity field, in which a fast spine is surrounded by a
slow layer. Note however that the slower jet component must be
relativistic in order to explain the anisotropic radiation of radio
galaxy cores (e.g. Capetti \& Celotti 1998).  The observed flux
is dominated by the emission from either the spine or the slower
layer, in the case of aligned and misaligned objects, respectively.
Interestingly, the existence of velocity structures in the jet has
been suggested by various authors (e.g. Laing 1999)
in order to explain some radio structures observed in FR~I
jets. Recently, observational evidence for such a behaviour has been
found in 3C~353 (an FR~II), M~87, B2~1144+35.
We thus considered this hypothesis in the simplest case, i.e. a model with 
two axisymmetric components having the same intrinsic luminosity, spectrum
and emitting volume (see Chiaberge et al. 2000 for details).
We found that Lorentz factors of the layer $\Gamma_{\rm layer}\sim 2$ 
can account for the unification of FR~I (of the 3CR) with LBL and intermediate
luminosity BL Lacs.  Instead the debeaming trails for the lowest
luminosity HBL do not cross the FR~I region in the $L_o-L_r$
plane. While the HBL behavior should be compared with that of radio
galaxies with which they share the extended radio power (e.g. those of
the B2 catalogue), our simple two-component jet model would not
account for the observed properties if the cores of such low-power
FR~I radio galaxies lied on the extrapolation of the 3CR radio-optical
correlation.  A possible interpretation is that in such 
weak sources the radio emitting region is less beamed than the
optical one, as could be expected if the jet decelerates after the
higher energy emitting zone.


\end{document}